\documentclass[aps,prb,reprint,superscriptaddress,longbibliography]{revtex4-2}
\usepackage{graphicx}
\usepackage{amsmath}
\usepackage{amsfonts}
\usepackage{bm}
\usepackage{cancel}
\usepackage[dvipsnames]{xcolor}
\usepackage[unicode=true, colorlinks=true, citecolor={blue!80!black}, urlcolor={blue!50!black}, linkcolor = {blue!80!black}]{hyperref}
\usepackage{microtype}
\usepackage[normalem]{ulem}
\usepackage{mathrsfs}
\usepackage{multirow}
\usepackage{array}
\usepackage{lipsum}
\usepackage{mathtools}

\DeclarePairedDelimiter\floor{\lfloor}{\rfloor}

\newcommand{\abs}[1]{\left\vert#1\right\vert}
\newcommand{\ket}[1]{\left\vert#1\right\rangle}
\newcommand{\bra}[1]{\left\langle#1\right\vert}

\newcommand{\up}{\uparrow}
\newcommand{\down}{\downarrow}

\graphicspath{{../figures/}}

\begin{document}

\title{
Andreev spin qubit protected by Franck-Condon blockade
}

\author{P. D. Kurilovich}
\affiliation{Departments of Applied Physics and Physics, Yale University, New Haven, CT 06520, USA}
\author{T. Vakhtel}
\affiliation{QuTech and Kavli Institute of Nanoscience, Delft University of Technology, Delft 2628 CJ, The Netherlands}
\author{T.~Connolly}
\affiliation{Departments of Applied Physics and Physics, Yale University, New Haven, CT 06520, USA}
\author{C.~G.~L.~B\o ttcher}
\affiliation{Departments of Applied Physics and Physics, Yale University, New Haven, CT 06520, USA}
\author{B. van Heck}
\affiliation{Dipartimento di Fisica, Sapienza Università di Roma, Piazzale Aldo Moro 2, 00185 Rome, Italy}

\date{\today}
\begin{abstract}
Andreev levels localized in a weak link between two superconductors can trap a superconducting quasiparticle. If there is a spin-orbit coupling in the link, the spin of the quasiparticle couples to the
Josephson current. This effect can be leveraged to control and readout the spin of the quasiparticle thus using it as a qubit. One of the factors limiting the performance of such an Andreev spin qubit is spin relaxation. Here, we theoretically demonstrate that the relaxation lifetime can be enhanced by utilizing the coupling between the Andreev spin and the supercurrent in a transmon circuit. The coupling ensures that the flip of the quasiparticle spin can only happen if it is accompanied by the excitation of multiple plasmons, as dictated by the Franck-Condon principle. This blocks spin relaxation at temperatures small compared to plasmon energy.
\end{abstract}

\maketitle

\section{Introduction}

The spin of a single quasiparticle trapped in a Josephson junction can couple to the supercurrent via the spin-orbit interaction~\cite{chtchelkatchev2003}.
Following several years of theoretical work~\cite{chtchelkatchev2003,beri2006,dellanna2008,michelsen2008,padurariu2010,reynoso2012,park2017}, this effect has been experimentally established via microwave spectroscopy measurements~\cite{tosi2019,metzger2021}.
It is a key element for the operation of Andreev spin qubits~\cite{hays2021,pita2023,pita2024}. 
These novel qubits have multiple properties favorable for quantum computing: a $\mu\mathrm{m}$-scale spatial footprint, large relative anharmonicity, and a direct interaction with microwave fields mediated by the spin-supercurrent coupling. The latter coupling also allows for cm-scale qubit connectivity~\cite{pita2024} which is useful for quantum simulations~\cite{pita2025}.

In the current experimental realizations of Andreev spin qubits, however, the benefits listed above are outweighed by short decoherence times~\cite{pita2023,lu2025,hoffman2025}. Both dephasing and relaxation times of Andreev spin qubits are much shorter than those of state-of-the-art semiconductor spin qubits~\cite{stano2022,burkard2023}.
So far, all measured devices were based on InAs/Al nanowires, which are far from optimal for spin coherence~\cite{stano2022}, in large part due to the presence of a nuclear spin bath.

A strategy to increase the dephasing time is to update the material platform used in Andreev spin qubit devices to one with less nuclear spins. A promising example is isotopically purified germanium~\cite{sigillito2015}.
Notably, Ge-based quantum dot Josephson junctions have been realized~\cite{lakic2025}, and theoretical proposals for Andreev spin qubits in the material have been developed~\cite{hoffman2025ge}.

In contrast, the origin of relaxation in Andreev spin qubits is still under active investigation~\cite{lu2025}, and therefore a strategy to improve relaxation time is still lacking.
Material considerations aside, one may wonder: in the spirit of hardware-protected qubits~\cite{gyenis2021}, is it possible to enhance the relaxation time of the Andreev spin qubit by circuit design?

\begin{figure}[t!]
    \includegraphics{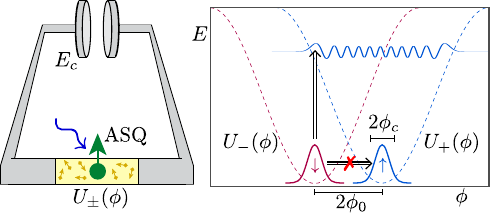}
    \caption{Protected Andreev spin qubit. \emph{Left:} A spin-$1/2$ quasiparticle (green) trapped inside a Josephson junction forming a resonant transmon circuit. The quasiparticle spin couples to the supercurrent flowing in the circuit via the spin-orbit coupling, leading to a spin-dependent Josephson potential $U_\pm(\phi)$. The noise in the magnetic environment of the spin (dark yellow spins) causes it to relax.
    The quasiparticle spin can also be externally driven (blue arrow), e.g. via a local gate. \emph{Right:} The potential energy of the circuit splits in two branches (dashed lines), one for spin up (blue) and one for spin down (red).
    The transmon ground state wave functions for opposite spins are disjoint, which protects the spin from relaxation.
    The protection can be compromised by a finite temperature, which activates transitions where circuit plasmons are excited in addition to the spin flip (vertical arrow).
    By the Franck-Condon principle, the matrix elements for such transitions are larger than those without plasmons (horizontal arrow).}
    \label{fig:setup}
\end{figure}

In this work, we show that the relaxation time of the Andreev spin qubit can be enhanced by shunting it with a capacitor, as in a transmon circuit~\cite{koch2007}.
Then, provided the spin-dependent supercurrent is large enough, the transmon ground state wave-functions for opposite spin orientations become disjoint, separated by a distance $\sim2\phi_0$ larger than their extent $\sim2\phi_c$, see Fig.~\ref{fig:setup}.
In this situation, the matrix elements for spin-flip transitions are exponentially suppressed with $(\phi_0/\phi_c)^2$, leading to an enhancement of the spin relaxation time.
The suppression of transition matrix elements is governed by Franck-Condon principle~\cite{bransden2003}, and we therefore refer to it as the Franck-Condon blockade, by analogy with a related phenomenon in molecular transport~\cite{koch2005,leturcq2009}.
The pure dephasing time of the Andreev spin qubit is neither improved nor worsened by Franck-Condon blockade.

We build a systematic theory of Franck-Condon blockade of Andreev spin qubit relaxation.
We describe how the protection can be compromised by  finite temperature. In fact, we show that the residual spin relaxation comes from a \textit{thermally-activated process} where a spin-flip is accompanied by an excitation of several plasmonic quanta.
Despite requiring energy, this process has a larger matrix element compared to the excitation-less process, as described by the Franck-Condon principle.
Additionally, we identify experimentally observable consequences of the Franck-Condon blockaded regime: the appearance of plasmonic transitions in response to an external drive acting on the spin and a staircase-shaped dependence of spin lifetime on magnetic field.
Finally, we provide a guideline for future experiments by discussing the device design of the protected Andreev spin qubit.
Throughout the paper we use units with $k_B = \hbar = 1$.

\section{Model and main idea}

We consider a circuit in which the junction trapping the spin-$1/2$ quasiparticle is shunted by capacitance $C$, see Fig.~\ref{fig:setup}. Adopting a minimal model for the Josephson energy~\cite{padurariu2010}, the Hamiltonian of the circuit is given by
\begin{equation}\label{eq:toymodel}
H_0 = 4E_c (i\partial_\phi+n_g)^2 + E_0 \cos(\phi) + E_\textrm{so}\sigma_z\sin(\phi)\,.
\end{equation}
Here, $\phi$ is the superconducting phase difference, while $\sigma_{x,y,z}$ are spin Pauli matrices. $E_c=e^2/2C$ is the single-electron charging energy, and $n_g$ is a shift due to stray offset charges. The couplings $E_0$ and $E_\textrm{so}$ represent the spin-independent and spin-dependent contributions to the amplitude of Cooper pair tunneling across the junction in the presence of the trapped quasiparticle.
This model has been applied successfully to quantum dot Josephson junctions in InAs/Al devices~\cite{bargerbos2023}.

The last term of Eq.~\eqref{eq:toymodel} arises due to spin-orbit interaction.
It is the simplest coupling between spin and superconducting phase compatible with time-reversal symmetry (which sends $\phi\to-\phi$ and $\sigma_z\to-\sigma_z$) and the compactness of $\phi$.
In principle, the direction $\hat{n}$ of the spin-orbit field may depend on the phase difference, so that $\sigma_z$ should be replaced by a more general operator $\vec{\sigma}\cdot\hat{n}(\phi)$ with $\hat{n}(\phi)=\hat{n}(-\phi)$. This replacement opens new channels of spin relaxation, as it couples the spin to charge fluctuations via $n_g$. This, however, does not qualitatively affect our results regarding the Franck-Condon blockade. We therefore resort to the simple form of the coupling Hamiltonian given by Eq.~\eqref{eq:toymodel}.

The potential in Eq.~\eqref{eq:toymodel} consists of two sinusoidal Josephson energy branches of equal amplitude $\sqrt{E_0^2+E_\textrm{so}^2}$, see Fig.~\ref{fig:setup}. The energy minima occur at $\phi = \pi \pm \phi_0$ for spin up and down states respectively, with
\begin{equation}\label{eq:phi0}
\phi_0 = \arctan(E_\textrm{so}/E_0)\,.
\end{equation}
The size of the quantum fluctuations is set by
\begin{equation}\label{eq:phic}
\phi_c = \sqrt{\frac{4 E_c}{\omega_p}}\,,
\end{equation}
where $\omega_p=\sqrt{8E_c (E_0^2+E_\textrm{so}^2)^{1/2}}$ is the Josephson plasma frequency of the small oscillations of the phase around the minima. Each energy level in the system is \emph{Kramers-degenerate} due to the presence of the spin degree of freedom.
When $E_c\ll\omega_p$, the Josephson potential can be approximated by a parabola and the low-lying levels are equidistant, $E_k=\omega_p(k+\tfrac{1}{2})$, where the integer number $k$ is the number of plasmon excitations. In this limit, one can neglect the exponentially small dependence of energy levels on $n_g$. In a more accurate treatment, which is beyond the scope of the manuscript, the $E_k$ are the energy levels of the Cooper-pair box Hamiltonian~\cite{koch2007}.

To describe Franck-Condon blockade, we introduce a parameter $\xi_0=(\phi_0/\phi_c)^2$.
If $\xi_0~ \gg~1$, the lowest-level wave functions with opposite spin are disjoint.
Their overlap is exponentially small in $\xi_0$, since it is the integral of a product of two Gaussian wave functions of width $\sim \phi_c$ at a distance $\sim \phi_0$, see Fig.~\ref{fig:setup}.
We call regime $\xi_0\gg 1$ ``the strong coupling regime'', referring to the strength of coupling between the spin and the supercurrent.
To achieve the strong coupling regime it is sufficient to fulfill two conditions: (i) $\omega_p\gg E_c$ such that the circuit is in the transmon regime~\cite{koch2007} with small zero-point phase fluctuations ($\phi_c\ll 1$), (ii) $E_\textrm{so}\gtrsim E_0$ such that the separation between potential minima for opposite spin is large ($\phi_0 \sim 1$).

\begin{figure*}[t!]
    \includegraphics[width=0.9\textwidth]{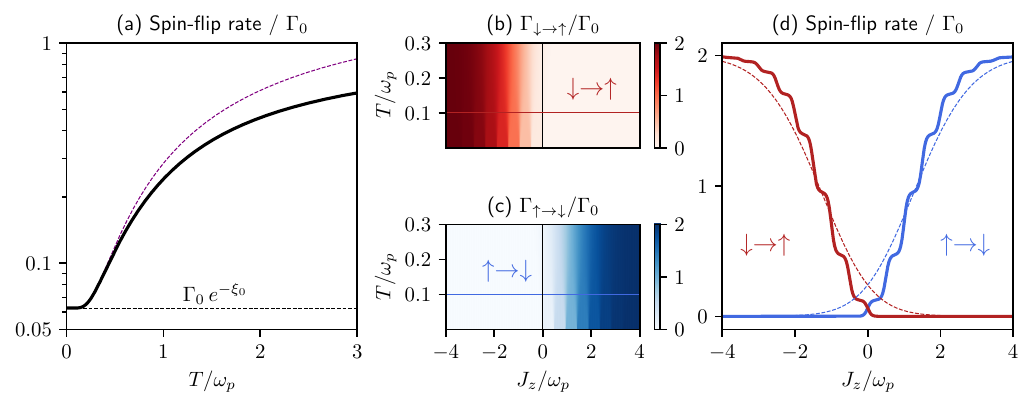}
    \caption{\emph{(a):} Temperature dependence of the spin-flip rate at $J_z=0$. We assume that the system is initialized in the plasmon ground state $k=0$ for one of the two spins. The solid line is Eq.~\eqref{eq:rate_sum}, the purple dashed line is the approximate expression~\eqref{eq:gamma_vs_T}, and the horizontal dashed line is the $T=0$ result~\eqref{eq:simple_protection}. The parameters are $E_c/E_\textrm{so}=0.035$ and $E_0/E_\textrm{so}=1$, yielding $\omega_p/E_\textrm{so}\approx 0.63$ and, using Eq.~\eqref{eq:phi0} and~\eqref{eq:phic}, $\xi_0\approx 2.8$. \emph{(b)-(c):} Dependence of the spin-flip rate on temperature and magnetic field parallel to the spin-orbit axis, starting from either spin down (b) or spin up (c). The rates display step-like features as a function of the magnetic field, which are washed out by increasing temperature. The vertical and horizontal lines correspond to the line cuts shown in panels (a) and (d) respectively. \emph{(d):} Magnetic field dependence of the spin-flip rate for initial states with opposite spin (initialized in a state with $k=0$). The solid curves are computed at $T/\omega_p = 0.1$, while the dashed ones at $T/\omega_p = 1$. In the low temperature curves, the steps occur at half-integer values of $J_z/\omega_p$, and have heights given by the coefficients $w_k$. The results presented in this figure are derived for a simple bath of two-level systems, with a constant density of states and energy-independent coupling to the Andreev spin.}
    \label{fig:spin_relaxation_rate}
\end{figure*}

Assume now that the Andreev spin is coupled to a vector field $\vec{J}$. This vector field can describe either an intentionally applied magnetic field or the uncontrolled noise stemming from the environment of the spin. 
To demonstrate the essence of the Franck-Condon blockade, we can momentarily leave the origin of $\vec{J}$ unspecified.

The coupling between the Andreev spin qubit and $\vec{J}$ can be incorporated into the model by adding an extra term 
\begin{equation}
H_J = \vec{J}\cdot\vec{\sigma}
\end{equation}
to the Hamiltonian $H_0$ of Eq.~\eqref{eq:toymodel}. Expressed in the basis of eigenstates of $H_0$, labeled by the discrete quantum number $k$ and the spin quantum number $\sigma=\up,\down$, the total Hamiltonian $H=H_0+H_J$ reads
\begin{align}\nonumber
H &= \sum_{k,\sigma} E_k\,\ket{k,\sigma}\bra{k,\sigma} \\
&+ \vec{J}\,\cdot\,\sum_{kk'}\sum_{\sigma,\sigma'}\,\bra{k,\sigma}\vec{\sigma}\ket{k',\sigma'}\ket{k,\sigma}\bra{k',\sigma'}\,.
\end{align}
We emphasize again that the unperturbed energy levels $E_k$ are Kramers degenerate and have wave functions centered around $\pi\pm\phi_0$, depending on $\sigma$. This relative shift decreases the overlap of wave functions with opposite spin $\sigma$ and therefore suppresses the matrix elements $\bra{k,\sigma}\vec{\sigma}\ket{k',\sigma'}$ if $\sigma\neq \sigma'$.

As mentioned above, in the harmonic limit $\omega_p\gg E_c$, the quantum number $k$ identifies the number of plasmon excitations and the wave functions are those of the harmonic oscillator. In this limit, if we project $H$ onto the Kramers-degenerate ground state space, that is onto the $k=0$ subspace, we obtain the effective Andreev spin Hamiltonian
\begin{equation}\label{eq:Heff}
H_\textrm{eff} = E_0 + e^{-\xi_0/2}(J_x\sigma_x+J_y\sigma_y) + J_z \sigma_z\,.
\end{equation}
We see that the matrix elements between ground states of opposite spins are suppressed a factor $e^{-\xi_0/2}$ which is small when $\xi_0\gg 1$. Thus, the external field is not effective in flipping the Andreev spin, and the qubit encoded in this spin is protected from relaxation. This is the main idea of our proposal.

\section{Spin relaxation rate}

We now show that Eq.~\eqref{eq:Heff} leads to the suppression of spin relaxation. To this end, we assume that the field ${\vec{J}}$ encodes a spin-flipping environment of the Andreev spin.
Realistically, such environment will be constituted by nuclear spins in combination with phonons \cite{erlingsson2002}, exchanged-coupled itinerant quasiparticles in the superconducting leads \cite{glazman2021}, magnetic field noise, and charge noise. However, on a qualitative level, the nature of the environment is not important for the discussion of Franck-Condon blockade. Therefore, to be concrete, we adopt a simplified model in which the environment is a collection of two-level systems of an unspecified origin.

The Hamiltonian of this simplified model is
\begin{align}
H_\textrm{tot} &= H_0 + \sum_i\,\epsilon_i \sigma_{z,i} + \sigma_x\,J_x\,,\\
J_x&=g\sum_i\sigma_{x,i}\,,
\end{align}
where $H_0$ is the Andreev spin qubit Hamiltonian of Eq.~\eqref{eq:toymodel}, $i$ runs over all the two level systems in the environment, and $\epsilon_i$ is the energy associated with each two level system.
We assume for simplicity that the energies are uniformly distributed, so that the environment has a constant density of states $\nu_0$.
The last term describes the coupling between the Andreev spin and the two level systems, and has been chosen in a form that allows Andreev spin flips induced by the environment.
Again for simplicity, we have assumed that the coupling constant $g$ in the last term does not depend on $i$. These simplifying assumptions can be relaxed without qualitative consequences for the results that follow.

We can compute the spin relaxation rate $\Gamma_{\uparrow \rightarrow \downarrow}$ as the transition rate between the two degenerate ground states of $H_0$. Within this simple bath model, using Fermi's golden rule, we find 
\begin{equation}   \label{eq:simple_protection}
\Gamma_{\uparrow \rightarrow \downarrow} = \Gamma_0 e^{-\xi_0},\quad \Gamma_0 = 2\pi \nu_0 |g|^2.
\end{equation}
As $\xi_0$ is increased, the spin relaxation time increases exponentially. We now explore limitations to this result.

\subsection{Temperature dependence}

The effective Hamiltonian given by Eq.~\eqref{eq:Heff} does not take into account the plasmonic excitations, that is all the states with $k>0$. At the same time, the full version of the coupling Hamiltonian $H_J$ mixes the low-energy states, $k=0$, with the excited states, $k>0$. For this reason, when describing the dynamics of the Andreev spin, neglecting the excited states is only allowed when the temperature is smaller than $\omega_p$. Otherwise, a more careful calculation of the spin-flip dynamics is warranted. In fact, as we now demonstrate, a temperature as low as $\omega_p / \ln(\xi_0)$ can significantly compromise the protection.

To reveal the role of finite temperature, we resort to the same two-level system model for the bath described above. For simplicity, we assume that the system is initialized in the plasmon ground state, $k=0$, for one of the spin directions. The total spin-flip rate can then be evaluated as
\begin{equation}
    \label{eq:rate_sum}
    \Gamma_{\uparrow \rightarrow \downarrow} = 2\Gamma_0 \sum_{k} \frac{w_k}{1 + e^{\beta \omega_p k}}\,,\;\; w_k=\frac{e^{-\xi_0} \xi_0^k}{k!}\,,
\end{equation}
with $\beta=1/T$. The weights $w_k\equiv \abs{g_{k0}}^2$ originate from the Franck-Condon overlap factor $g_{kk^\prime}$, defined as
\begin{equation}\label{eq:fc}
g_{kk'}=  \bra{k\!\downarrow} \sigma_x\ket{k^\prime\! \uparrow} = \int_{-\infty}^{\infty} \psi_k(\phi+\phi_0)\, \psi_{k'}(\phi-\phi_0)d\phi\,, 
\end{equation}
where $\psi_k$ are the wave functions of the harmonic oscillator.
At small temperature $T\ll\omega_p$, factors $(1+e^{\beta\omega_p k})^{-1}$ in Eq.~\eqref{eq:rate_sum} can be approximated as $e^{-\beta \omega_p k}\ll 1$ for any $k>0$. Then, we find
\begin{equation}\label{eq:gamma_vs_T}
    \Gamma_{\uparrow\rightarrow\downarrow} = \Gamma_0 e^{-\xi_0} \left(2 e^{\xi_0 e^{-\beta\omega_p}} - 1\right).
\end{equation}
According to this equation, Eq.~\eqref{eq:simple_protection} is accurate as long as the temperature satisfies $T\ll \omega_p / \ln \xi_0$. Otherwise, Eq.~\eqref{eq:simple_protection} underestimates the transition rate. The rate is enhanced due to the possibility of a spin-flip with a simultaneous excitation of $k$ plasmons. Even though this process is penalized by a factor $e^{-\beta\omega_pk}$, it has a much larger matrix element compared to that for a process unaccompanied by plasmon excitations. In the regime $T \gg \omega_p$, we obtain $\Gamma_{\uparrow\rightarrow \downarrow} = \Gamma_0$ since $e^{\beta \omega_p k}\approx 1$ for all $k$ and the following sum rule holds:
\begin{equation}\label{eq:sum_rule}
\sum_k w_k = e^{-\xi_0} \sum_k \frac{\xi_0^{k}}{k!} = 1\,.
\end{equation}
The temperature dependence of the spin relaxation rate is illustrated in Fig.~\ref{fig:spin_relaxation_rate}a, which includes a comparison between Eq.~\eqref{eq:rate_sum} and its approximation Eq.~\eqref{eq:gamma_vs_T}.

Finally, we note that Eq.~\eqref{eq:rate_sum} can be extended to the case in which the plasmons are thermally occupied (as opposed to taking $k=0$ as the initial state). This brings no qualitative change to the conclusions of this section. 

\section{Experimental signatures}

We now discuss two experimental signatures of a protected regime of the Andreev spin qubit for $T\ll \omega_p$. The first such signature is the unusual magnetic field dependence of the spin-flip rate. Another is the appearance of plasmonic branches in the spectrum of a driven Andreev spin qubit.

\subsection{Effect of a static magnetic field}

To see this dependence, we assume that a static magnetic field is applied (anti-)parallel to the spin-orbit axis and so we include a term $J_z \sigma_z$ in the Hamiltonian.
We note that the magnitude and sign of $J_z$ are determined by the external magnetic field through the $g$-factor of the Andreev spin, which may depend on the material platform and the geometry of the junction~\cite{pita2023,lakic2025}.
Then, Eq.~\eqref{eq:rate_sum} generalizes to
\begin{equation}
    \label{eq:rate_sum_finite_B}
    \Gamma_{\uparrow \rightarrow \downarrow} = 2\Gamma_0 \sum_{k} \frac{w_k}{1 + e^{\beta \omega_p k}e^{-2\beta J_z}}.
\end{equation}
From this equation, we see that for $J_z<0$ and $T\lesssim |J_z|$, the spin-flip rate is additionally suppressed compared to Eq.~\eqref{eq:simple_protection}. Indeed, an extra energy of $2J_z$ is required to excite the initial spin-up state into a spin-down state, even without exciting plasmons.
For positive $J_z$, instead, an increasing magnetic field sequentially lifts the energy of the $\ket{0\uparrow}$ state above that of the $\ket{k\down}$ state. Each time this happens, a new relaxation channel opens up for the spin-up state. Hence, the spin-flip rate increases in steps~\cite{Frattini2024} centered around the points $J_z=k\omega_p/2$, see Fig.~\ref{fig:spin_relaxation_rate}d. The $k$-th step has height $\Gamma_0 w_k$.
The relaxation rate with an opposite initial spin has an inverted field dependence, $\Gamma_{\down\to\up}(J_z)=\Gamma_{\up\to\down}(-J_z)$, as illustrated in Fig.~\ref{fig:spin_relaxation_rate}b-c.

The Franck-Condon factors of Eq.~\eqref{eq:fc} also determine the splitting of Kramers-degenerate excited states due to magnetic field \emph{perpendicular} to the spin-orbit axis. Indeed, according to first-order perturbation theory, the term $H_J=J_x \sigma_x$ in the Hamiltonian splits the energy of the $k$-th level by $\delta E_k \approx J_x g_{kk} = J_x e^{-\xi_0/2} L_k(\xi_0)$, with $L_k$ the $k$-th Laguerre polynomial. 

\subsection{Driven excitation of plasmonic branches}

\begin{figure}[t!]
\includegraphics[width=\columnwidth]{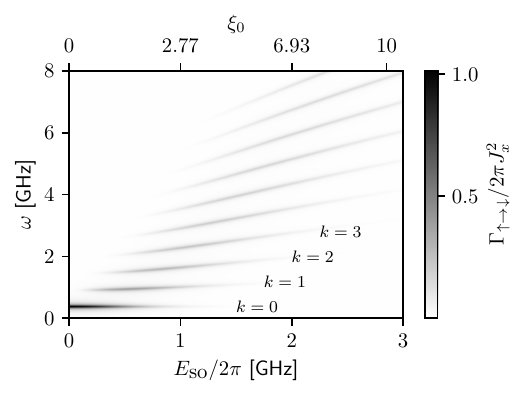}
    \caption{Spin-flip rate in response to an external drive at frequency $\omega$ as a function of the coupling $E_\textrm{so}$, using Eq.~\eqref{eq:drive_response}. We set $T=0$, $J_z/2\pi=0.2$~GHz, $E_0/2\pi=1$~GHz, $E_c/2\pi=35$~MHz, $\Gamma/2\pi=50$~MHz. The dark lines visible in the figure correspond to distinct spin-flip transitions, each accompanied by the simultaneous excitations of $k$ plasmons, with increasing values $k=0,1,2,3,\dots$ as indicated. The visibility of each transition depends on $E_\textrm{so}$ via corresponding Franck-Condon factors $w_k$, as in Eq.~\eqref{eq:drive_response}. The results illustrate that in the presence of spin-orbit coupling, the spin-flip transition fans out into a set of plasmonic transitions with spacing $\omega_p$.}
    \label{fig:external_drive}
\end{figure}

Consider a driving field coupled to $\sigma_x$, $H_J = J_x\cos(\omega t) \sigma_x$. If the frequency of the drive matches $2J_z$ then Rabi oscillations develop between states $|0\uparrow\rangle$ and $|0\downarrow\rangle$. However, $2J_z$ is not the only frequency activating the spin-flip process. Namely, if the driving frequency is resonant with a plasmonic transition, $\omega=k\omega_p+2J_z$, the driving field will selectively induce coherent oscillations between the states $\ket{0\up}$ and $\ket{k\down}$, with a Rabi frequency
\begin{equation}\label{eq:Rabi}
\Omega_{k0} = J_x \abs{g_{k0}} = J_x e^{-\xi_0/2} \xi_0^{k/2}/\sqrt{k!}\,.
\end{equation}
The energy levels excited by the drive may be broadened, such that the Rabi oscillations are over-damped. In this case, the spectrum should be characterized with a rate of incoherent spin-flip transitions at different drive frequencies. According to Fermi's golden rule, the rate is given by
\begin{align}\label{eq:drive_response}
\Gamma_{\up\to\down} &= 2\pi J_x^2\,\sum_{k=0}^{\infty} \frac{w_k}{\pi}\frac{\Gamma}{(\omega-k\omega_p-2J_z)^2+\Gamma^2}\,,
\end{align}
where $\Gamma$ is a parameter describing the broadening. The resulting rate as a function of the spin-orbit parameter $E_\mathrm{so}$ is shown in Fig.~\ref{fig:external_drive}.

In the absence of spin-orbit coupling ($E_\textrm{so}=0$, and thus $\xi_0=0$), one has $w_0=1$ and $w_{k>0}=0$. Therefore, as is apparent from Fig.~\ref{fig:external_drive}, the only active transition is the spin-flip transition with $k=0$.
At finite $E_\textrm{so}$, instead, the absorption spectrum consists of a set of distinct transitions involving the excitation of $k$ plasmons.
Indeed, the sum rule of Eq.~\eqref{eq:sum_rule} guarantees that the exponential suppression of $w_0$ is compensated by a power-law increase in the strength of higher-frequency transitions.
If $\xi_0\gg 1$, using Stirling's approximation, one can see that the maximum transition strength occurs for $k= \floor{\xi_0}$. Therefore, in the strong coupling regime, it is easier to excite a spin-flip transition if it is accompanied by the excitation of multiple plasmons, because of a larger wave function overlap with such a final state.

Finally, we note that although we considered a purely magnetic drive, it is also possible to coherently drive the spin via a gate, using electron dipole spin resonance~\cite{nowack2007,nadj2010,golovach2006}.

\section{Experimental feasibility}

Demonstrating the Frack-Condon blockade requires the realization of a strong coupling regime, $\xi_0\gg 1$, which is within reach of current experiments. Indeed, measurements of Andreev spins in InAs nanowires routinely observe $E_0/2\pi \sim E_\textrm{so}/2\pi \sim 1~\mathrm{GHz}$~\cite{tosi2019,metzger2021,bargerbos2023,lu2025}. In that case, to reach strong coupling it is sufficient to construct a circuit with a small charging energy, $E_C /2\pi \ll 1~\mathrm{GHz}$. Transmon circuits with values of $E_c$ as low as $36~\mathrm{MHz}$ have been engineered~\cite{kurilovich2025}. For these values, $\xi_0 \approx 3$ can be expected. Potentially, $\xi_0$ can be further increased by finding material platforms leading to higher values of $E_\textrm{so}$. Fig.~\ref{fig:external_drive} shows that for the parameter values quoted in this paragraph the experimental signatures of the protected regime are accessible within the typical bandwidth ($\approx 1-10$ GHz) of a circuit QED experiment.

Another challenge is the requirement $T\ll \omega_p$, but we believe that this condition can also be met in the laboratory. For the parameters listed above, we expect $\omega_p/2\pi\sim 600~\mathrm{MHz}$, that is $\sim 30~\mathrm{mK}$, above the $\sim 10~\mathrm{mK}$ achievable by modern dilution refrigerators. The feasibility of operating low-frequency transmons with small thermal populations was experimentally demonstrated in Ref.~\cite{kurilovich2025}.

Two final requirements for the operation of the protected Andreev spin qubit concern the feasibility of qubit readout and gates. The spin can be read out dispersively by coupling the circuit to a readout resonator~\cite{hays2021, bargerbos2023}. To achieve such a spin readout, time-reversal symmetry has to be broken during the measurement, for instance by applying a magnetic field perpendicular to spin-orbit field or by threading magnetic flux through a SQUID loop~\footnote{See e.g. the Supplemental Materiak of Ref.~\cite{bargerbos2023} for a demontration of Andreev spin readout via a resonator.}. The realization of single-qubit gates is complicated by the suppressed overlap between the transmon ground state wave functions for different spins. Fortunately, the Rabi frequency scales with $\xi_0$ more mildly, as $e^{-\xi_0/2}$, than relaxation, as $e^{-\xi_0}$. 
We leave a detailed analysis of single and two-qubit gates for future investigation.

\section{Conclusions}

We have shown that an Andreev spin qubit can be protected against spin relaxation by shunting it with a large capacitance.
This provides a circuit-based solution, agnostic to the material platform, to the challenge of improving the relaxation time of Andreev spin qubits.
We provided a set of specific experimental predictions that distinguish the protected regime from an unprotected one. These include an unusual dependence of the spin relaxation rate on the magnetic field as well as the appearance of additional branches detectable in the two-tone spectrum of the system.

We have highlighted the limitations on the protected regime posed by finite temperature of the device.
This thermal fragility can be related to the Franck-Condon principle via the sum rule of Eq.~\eqref{eq:sum_rule}.
The exponential suppression of spin flips due to the disjoint support of the \textit{low-lying} plasmon states is accompanied by the simultaneous increase of the matrix elements involving the \textit{excited} plasmon states. At finite temperature, this diminishes the expected returns of being in a protected regime.
A similar trade-off could apply to any protected qubit based on a double-well energy landscape~\cite{gyenis2021,pistolesi2021}.

The protected Andreev spin qubit that we propose has noise strongly biased towards pure dephasing. Such a noise bias may be advantageous for quantum error correction~\cite{aliferis2008,tuckett2018,tuckett2019}.
The strategy of improving relaxation time has been very successful e.g.~in cat qubits~\cite{reglade2024quantum}.
This suggests that improving dephasing time, while beneficial, does not need to be the main bottleneck for the development of Andreev spin qubits.
Along these lines, it will be interesting to merge the ideas presented in our manuscript with those of the recent blueprint for Kramers-protected error correction in Andreev spin qubit arrays~\cite{lu2025kramers}.
Other extensions are the inclusion of a superinductor~\cite{manucharyan2009} into the circuit~\cite{olmoinprep} and the study of the Franck-Condon blockade in multi-terminal junctions~\cite{heck2014,matute2024,piasotski2024}.

\begin{acknowledgments}
We thank Valla Fatemi, Lukas Splitthoff, Spencer Diamond, Heekun Nho, Vladislav Kurilovich, Christian K. Andersen, Arno Bargerbos, Marta Pita-Vidal, Jaap Wesdorp, Maximilian Rimbach-Russ, Aleksandr Svetogorov, Leandro Tosi, Leonid Glazman, and Michel Devoret for useful discussions and feedback on the manuscript.

The research of P.D.K., T.C., and C.G.L.B.~was sponsored by the Army Research Office (ARO) under grants no.~W911NF-22-1- 0053 and W911NF-23-1-0051, by DARPA under grant no. HR0011-24-2-0346, by the U.S. Department of Energy (DoE), Office of Science, National Quantum Information Science Research Centers, Co-design Center for Quantum Advantage (C2QA) under contract number DE-SC0012704. The views and conclusions contained in this document are those of the authors and should not be interpreted as representing the official policies, either expressed or implied, of the ARO, DARPA, DoE, or the US Government. The US Government is authorized to reproduce and distribute reprints for Government purposes notwithstanding any copyright notation herein.
\end{acknowledgments}

\subsection*{Data availability}

The data that support the findings of this article are openly available~\cite{zenodo}.

\bibliography{references}

\begin{thebibliography}{46}%
\makeatletter
\providecommand \@ifxundefined [1]{%
 \@ifx{#1\undefined}
}%
\providecommand \@ifnum [1]{%
 \ifnum #1\expandafter \@firstoftwo
 \else \expandafter \@secondoftwo
 \fi
}%
\providecommand \@ifx [1]{%
 \ifx #1\expandafter \@firstoftwo
 \else \expandafter \@secondoftwo
 \fi
}%
\providecommand \natexlab [1]{#1}%
\providecommand \enquote  [1]{``#1''}%
\providecommand \bibnamefont  [1]{#1}%
\providecommand \bibfnamefont [1]{#1}%
\providecommand \citenamefont [1]{#1}%
\providecommand \href@noop [0]{\@secondoftwo}%
\providecommand \href [0]{\begingroup \@sanitize@url \@href}%
\providecommand \@href[1]{\@@startlink{#1}\@@href}%
\providecommand \@@href[1]{\endgroup#1\@@endlink}%
\providecommand \@sanitize@url [0]{\catcode `\\12\catcode `\$12\catcode `\&12\catcode `\#12\catcode `\^12\catcode `\_12\catcode `\%12\relax}%
\providecommand \@@startlink[1]{}%
\providecommand \@@endlink[0]{}%
\providecommand \url  [0]{\begingroup\@sanitize@url \@url }%
\providecommand \@url [1]{\endgroup\@href {#1}{\urlprefix }}%
\providecommand \urlprefix  [0]{URL }%
\providecommand \Eprint [0]{\href }%
\providecommand \doibase [0]{https://doi.org/}%
\providecommand \selectlanguage [0]{\@gobble}%
\providecommand \bibinfo  [0]{\@secondoftwo}%
\providecommand \bibfield  [0]{\@secondoftwo}%
\providecommand \translation [1]{[#1]}%
\providecommand \BibitemOpen [0]{}%
\providecommand \bibitemStop [0]{}%
\providecommand \bibitemNoStop [0]{.\EOS\space}%
\providecommand \EOS [0]{\spacefactor3000\relax}%
\providecommand \BibitemShut  [1]{\csname bibitem#1\endcsname}%
\let\auto@bib@innerbib\@empty
\bibitem [{\citenamefont {Chtchelkatchev}\ and\ \citenamefont {Nazarov}(2003)}]{chtchelkatchev2003}%
  \BibitemOpen
  \bibfield  {author} {\bibinfo {author} {\bibfnamefont {N.~M.}\ \bibnamefont {Chtchelkatchev}}\ and\ \bibinfo {author} {\bibfnamefont {Y.~V.}\ \bibnamefont {Nazarov}},\ }\bibfield  {title} {\bibinfo {title} {Andreev quantum dots for spin manipulation},\ }\href {https://doi.org/10.1103/PhysRevLett.90.226806} {\bibfield  {journal} {\bibinfo  {journal} {Phys. Rev. Lett.}\ }\textbf {\bibinfo {volume} {90}},\ \bibinfo {pages} {226806} (\bibinfo {year} {2003})}\BibitemShut {NoStop}%
\bibitem [{\citenamefont {B\'eri}\ \emph {et~al.}(2008)\citenamefont {B\'eri}, \citenamefont {Bardarson},\ and\ \citenamefont {Beenakker}}]{beri2006}%
  \BibitemOpen
  \bibfield  {author} {\bibinfo {author} {\bibfnamefont {B.}~\bibnamefont {B\'eri}}, \bibinfo {author} {\bibfnamefont {J.~H.}\ \bibnamefont {Bardarson}},\ and\ \bibinfo {author} {\bibfnamefont {C.~W.~J.}\ \bibnamefont {Beenakker}},\ }\bibfield  {title} {\bibinfo {title} {Splitting of {Andreev} levels in a {Josephson} junction by spin-orbit coupling},\ }\href {https://doi.org/10.1103/PhysRevB.77.045311} {\bibfield  {journal} {\bibinfo  {journal} {Phys. Rev. B}\ }\textbf {\bibinfo {volume} {77}},\ \bibinfo {pages} {045311} (\bibinfo {year} {2008})}\BibitemShut {NoStop}%
\bibitem [{\citenamefont {Dell'Anna}\ \emph {et~al.}(2007)\citenamefont {Dell'Anna}, \citenamefont {Zazunov}, \citenamefont {Egger},\ and\ \citenamefont {Martin}}]{dellanna2008}%
  \BibitemOpen
  \bibfield  {author} {\bibinfo {author} {\bibfnamefont {L.}~\bibnamefont {Dell'Anna}}, \bibinfo {author} {\bibfnamefont {A.}~\bibnamefont {Zazunov}}, \bibinfo {author} {\bibfnamefont {R.}~\bibnamefont {Egger}},\ and\ \bibinfo {author} {\bibfnamefont {T.}~\bibnamefont {Martin}},\ }\bibfield  {title} {\bibinfo {title} {Josephson current through a quantum dot with spin-orbit coupling},\ }\href {https://doi.org/10.1103/PhysRevB.75.085305} {\bibfield  {journal} {\bibinfo  {journal} {Phys. Rev. B}\ }\textbf {\bibinfo {volume} {75}},\ \bibinfo {pages} {085305} (\bibinfo {year} {2007})}\BibitemShut {NoStop}%
\bibitem [{\citenamefont {Michelsen}\ \emph {et~al.}(2008)\citenamefont {Michelsen}, \citenamefont {Shumeiko},\ and\ \citenamefont {Wendin}}]{michelsen2008}%
  \BibitemOpen
  \bibfield  {author} {\bibinfo {author} {\bibfnamefont {J.}~\bibnamefont {Michelsen}}, \bibinfo {author} {\bibfnamefont {V.~S.}\ \bibnamefont {Shumeiko}},\ and\ \bibinfo {author} {\bibfnamefont {G.}~\bibnamefont {Wendin}},\ }\bibfield  {title} {\bibinfo {title} {Manipulation with {Andreev} states in spin active mesoscopic {Josephson} junctions},\ }\href {https://doi.org/10.1103/PhysRevB.77.184506} {\bibfield  {journal} {\bibinfo  {journal} {Phys. Rev. B}\ }\textbf {\bibinfo {volume} {77}},\ \bibinfo {pages} {184506} (\bibinfo {year} {2008})}\BibitemShut {NoStop}%
\bibitem [{\citenamefont {Padurariu}\ and\ \citenamefont {Nazarov}(2010)}]{padurariu2010}%
  \BibitemOpen
  \bibfield  {author} {\bibinfo {author} {\bibfnamefont {C.}~\bibnamefont {Padurariu}}\ and\ \bibinfo {author} {\bibfnamefont {Y.~V.}\ \bibnamefont {Nazarov}},\ }\bibfield  {title} {\bibinfo {title} {Theoretical proposal for superconducting spin qubits},\ }\href {https://doi.org/10.1103/PhysRevB.81.144519} {\bibfield  {journal} {\bibinfo  {journal} {Phys. Rev. B}\ }\textbf {\bibinfo {volume} {81}},\ \bibinfo {pages} {144519} (\bibinfo {year} {2010})}\BibitemShut {NoStop}%
\bibitem [{\citenamefont {Reynoso}\ \emph {et~al.}(2012)\citenamefont {Reynoso}, \citenamefont {Usaj}, \citenamefont {Balseiro}, \citenamefont {Feinberg},\ and\ \citenamefont {Avignon}}]{reynoso2012}%
  \BibitemOpen
  \bibfield  {author} {\bibinfo {author} {\bibfnamefont {A.~A.}\ \bibnamefont {Reynoso}}, \bibinfo {author} {\bibfnamefont {G.}~\bibnamefont {Usaj}}, \bibinfo {author} {\bibfnamefont {C.~A.}\ \bibnamefont {Balseiro}}, \bibinfo {author} {\bibfnamefont {D.}~\bibnamefont {Feinberg}},\ and\ \bibinfo {author} {\bibfnamefont {M.}~\bibnamefont {Avignon}},\ }\bibfield  {title} {\bibinfo {title} {Spin-orbit-induced chirality of {Andreev} states in {Josephson} junctions},\ }\href {https://doi.org/10.1103/PhysRevB.86.214519} {\bibfield  {journal} {\bibinfo  {journal} {Phys. Rev. B}\ }\textbf {\bibinfo {volume} {86}},\ \bibinfo {pages} {214519} (\bibinfo {year} {2012})}\BibitemShut {NoStop}%
\bibitem [{\citenamefont {Park}\ and\ \citenamefont {Yeyati}(2017)}]{park2017}%
  \BibitemOpen
  \bibfield  {author} {\bibinfo {author} {\bibfnamefont {S.}~\bibnamefont {Park}}\ and\ \bibinfo {author} {\bibfnamefont {A.~L.}\ \bibnamefont {Yeyati}},\ }\bibfield  {title} {\bibinfo {title} {Andreev spin qubits in multichannel {Rashba} nanowires},\ }\href {https://doi.org/10.1103/PhysRevB.96.125416} {\bibfield  {journal} {\bibinfo  {journal} {Phys. Rev. B}\ }\textbf {\bibinfo {volume} {96}},\ \bibinfo {pages} {125416} (\bibinfo {year} {2017})}\BibitemShut {NoStop}%
\bibitem [{\citenamefont {Tosi}\ \emph {et~al.}(2019)\citenamefont {Tosi}, \citenamefont {Metzger}, \citenamefont {Goffman}, \citenamefont {Urbina}, \citenamefont {Pothier}, \citenamefont {Park}, \citenamefont {Yeyati}, \citenamefont {Nyg\aa{}rd},\ and\ \citenamefont {Krogstrup}}]{tosi2019}%
  \BibitemOpen
  \bibfield  {author} {\bibinfo {author} {\bibfnamefont {L.}~\bibnamefont {Tosi}}, \bibinfo {author} {\bibfnamefont {C.}~\bibnamefont {Metzger}}, \bibinfo {author} {\bibfnamefont {M.~F.}\ \bibnamefont {Goffman}}, \bibinfo {author} {\bibfnamefont {C.}~\bibnamefont {Urbina}}, \bibinfo {author} {\bibfnamefont {H.}~\bibnamefont {Pothier}}, \bibinfo {author} {\bibfnamefont {S.}~\bibnamefont {Park}}, \bibinfo {author} {\bibfnamefont {A.~L.}\ \bibnamefont {Yeyati}}, \bibinfo {author} {\bibfnamefont {J.}~\bibnamefont {Nyg\aa{}rd}},\ and\ \bibinfo {author} {\bibfnamefont {P.}~\bibnamefont {Krogstrup}},\ }\bibfield  {title} {\bibinfo {title} {Spin-orbit splitting of {Andreev} states revealed by microwave spectroscopy},\ }\href {https://doi.org/10.1103/PhysRevX.9.011010} {\bibfield  {journal} {\bibinfo  {journal} {Phys. Rev. X}\ }\textbf {\bibinfo {volume} {9}},\ \bibinfo {pages} {011010} (\bibinfo {year} {2019})}\BibitemShut {NoStop}%
\bibitem [{\citenamefont {Metzger}\ \emph {et~al.}(2021)\citenamefont {Metzger}, \citenamefont {Park}, \citenamefont {Tosi}, \citenamefont {Janvier}, \citenamefont {Reynoso}, \citenamefont {Goffman}, \citenamefont {Urbina}, \citenamefont {Levy~Yeyati},\ and\ \citenamefont {Pothier}}]{metzger2021}%
  \BibitemOpen
  \bibfield  {author} {\bibinfo {author} {\bibfnamefont {C.}~\bibnamefont {Metzger}}, \bibinfo {author} {\bibfnamefont {S.}~\bibnamefont {Park}}, \bibinfo {author} {\bibfnamefont {L.}~\bibnamefont {Tosi}}, \bibinfo {author} {\bibfnamefont {C.}~\bibnamefont {Janvier}}, \bibinfo {author} {\bibfnamefont {A.~A.}\ \bibnamefont {Reynoso}}, \bibinfo {author} {\bibfnamefont {M.~F.}\ \bibnamefont {Goffman}}, \bibinfo {author} {\bibfnamefont {C.}~\bibnamefont {Urbina}}, \bibinfo {author} {\bibfnamefont {A.}~\bibnamefont {Levy~Yeyati}},\ and\ \bibinfo {author} {\bibfnamefont {H.}~\bibnamefont {Pothier}},\ }\bibfield  {title} {\bibinfo {title} {{Circuit-QED} with phase-biased {Josephson} weak links},\ }\href {https://doi.org/10.1103/PhysRevResearch.3.013036} {\bibfield  {journal} {\bibinfo  {journal} {Phys. Rev. Res.}\ }\textbf {\bibinfo {volume} {3}},\ \bibinfo {pages} {013036} (\bibinfo {year} {2021})}\BibitemShut {NoStop}%
\bibitem [{\citenamefont {Hays}\ \emph {et~al.}(2021)\citenamefont {Hays}, \citenamefont {Fatemi}, \citenamefont {Bouman}, \citenamefont {Cerrillo}, \citenamefont {Diamond}, \citenamefont {Serniak}, \citenamefont {Connolly}, \citenamefont {Krogstrup}, \citenamefont {Nygård}, \citenamefont {Yeyati}, \citenamefont {Geresdi},\ and\ \citenamefont {Devoret}}]{hays2021}%
  \BibitemOpen
  \bibfield  {author} {\bibinfo {author} {\bibfnamefont {M.}~\bibnamefont {Hays}}, \bibinfo {author} {\bibfnamefont {V.}~\bibnamefont {Fatemi}}, \bibinfo {author} {\bibfnamefont {D.}~\bibnamefont {Bouman}}, \bibinfo {author} {\bibfnamefont {J.}~\bibnamefont {Cerrillo}}, \bibinfo {author} {\bibfnamefont {S.}~\bibnamefont {Diamond}}, \bibinfo {author} {\bibfnamefont {K.}~\bibnamefont {Serniak}}, \bibinfo {author} {\bibfnamefont {T.}~\bibnamefont {Connolly}}, \bibinfo {author} {\bibfnamefont {P.}~\bibnamefont {Krogstrup}}, \bibinfo {author} {\bibfnamefont {J.}~\bibnamefont {Nygård}}, \bibinfo {author} {\bibfnamefont {A.~L.}\ \bibnamefont {Yeyati}}, \bibinfo {author} {\bibfnamefont {A.}~\bibnamefont {Geresdi}},\ and\ \bibinfo {author} {\bibfnamefont {M.~H.}\ \bibnamefont {Devoret}},\ }\bibfield  {title} {\bibinfo {title} {Coherent manipulation of an {Andreev} spin qubit},\ }\href {https://doi.org/10.1126/science.abf0345} {\bibfield  {journal} {\bibinfo  {journal} {Science}\ }\textbf {\bibinfo {volume} {373}},\
  \bibinfo {pages} {430} (\bibinfo {year} {2021})}\BibitemShut {NoStop}%
\bibitem [{\citenamefont {Pita-Vidal}\ \emph {et~al.}(2023)\citenamefont {Pita-Vidal}, \citenamefont {Bargerbos}, \citenamefont {{\v{Z}}itko}, \citenamefont {Splitthoff}, \citenamefont {Gr{\"u}nhaupt}, \citenamefont {Wesdorp}, \citenamefont {Liu}, \citenamefont {Kouwenhoven}, \citenamefont {Aguado}, \citenamefont {van Heck} \emph {et~al.}}]{pita2023}%
  \BibitemOpen
  \bibfield  {author} {\bibinfo {author} {\bibfnamefont {M.}~\bibnamefont {Pita-Vidal}}, \bibinfo {author} {\bibfnamefont {A.}~\bibnamefont {Bargerbos}}, \bibinfo {author} {\bibfnamefont {R.}~\bibnamefont {{\v{Z}}itko}}, \bibinfo {author} {\bibfnamefont {L.~J.}\ \bibnamefont {Splitthoff}}, \bibinfo {author} {\bibfnamefont {L.}~\bibnamefont {Gr{\"u}nhaupt}}, \bibinfo {author} {\bibfnamefont {J.~J.}\ \bibnamefont {Wesdorp}}, \bibinfo {author} {\bibfnamefont {Y.}~\bibnamefont {Liu}}, \bibinfo {author} {\bibfnamefont {L.~P.}\ \bibnamefont {Kouwenhoven}}, \bibinfo {author} {\bibfnamefont {R.}~\bibnamefont {Aguado}}, \bibinfo {author} {\bibfnamefont {B.}~\bibnamefont {van Heck}}, \emph {et~al.},\ }\bibfield  {title} {\bibinfo {title} {Direct manipulation of a superconducting spin qubit strongly coupled to a transmon qubit},\ }\href {https://doi.org/https://doi.org/10.1038/s41567-023-02071-x} {\bibfield  {journal} {\bibinfo  {journal} {Nature Physics}\ }\textbf {\bibinfo {volume} {19}},\ \bibinfo {pages} {1110}
  (\bibinfo {year} {2023})}\BibitemShut {NoStop}%
\bibitem [{\citenamefont {Pita-Vidal}\ \emph {et~al.}(2024)\citenamefont {Pita-Vidal}, \citenamefont {Wesdorp}, \citenamefont {Splitthoff}, \citenamefont {Bargerbos}, \citenamefont {Liu}, \citenamefont {Kouwenhoven},\ and\ \citenamefont {Andersen}}]{pita2024}%
  \BibitemOpen
  \bibfield  {author} {\bibinfo {author} {\bibfnamefont {M.}~\bibnamefont {Pita-Vidal}}, \bibinfo {author} {\bibfnamefont {J.~J.}\ \bibnamefont {Wesdorp}}, \bibinfo {author} {\bibfnamefont {L.~J.}\ \bibnamefont {Splitthoff}}, \bibinfo {author} {\bibfnamefont {A.}~\bibnamefont {Bargerbos}}, \bibinfo {author} {\bibfnamefont {Y.}~\bibnamefont {Liu}}, \bibinfo {author} {\bibfnamefont {L.~P.}\ \bibnamefont {Kouwenhoven}},\ and\ \bibinfo {author} {\bibfnamefont {C.~K.}\ \bibnamefont {Andersen}},\ }\bibfield  {title} {\bibinfo {title} {Strong tunable coupling between two distant superconducting spin qubits},\ }\href {https://doi.org/https://doi.org/10.1038/s41567-024-02497-x} {\bibfield  {journal} {\bibinfo  {journal} {Nature Physics}\ ,\ \bibinfo {pages} {1}} (\bibinfo {year} {2024})}\BibitemShut {NoStop}%
\bibitem [{\citenamefont {Pita-Vidal}\ \emph {et~al.}(2025)\citenamefont {Pita-Vidal}, \citenamefont {Wesdorp},\ and\ \citenamefont {Andersen}}]{pita2025}%
  \BibitemOpen
  \bibfield  {author} {\bibinfo {author} {\bibfnamefont {M.}~\bibnamefont {Pita-Vidal}}, \bibinfo {author} {\bibfnamefont {J.~J.}\ \bibnamefont {Wesdorp}},\ and\ \bibinfo {author} {\bibfnamefont {C.~K.}\ \bibnamefont {Andersen}},\ }\bibfield  {title} {\bibinfo {title} {Blueprint for all-to-all-connected superconducting spin qubits},\ }\href {https://doi.org/10.1103/PRXQuantum.6.010308} {\bibfield  {journal} {\bibinfo  {journal} {PRX Quantum}\ }\textbf {\bibinfo {volume} {6}},\ \bibinfo {pages} {010308} (\bibinfo {year} {2025})}\BibitemShut {NoStop}%
\bibitem [{\citenamefont {Lu}\ \emph {et~al.}(2025{\natexlab{a}})\citenamefont {Lu}, \citenamefont {Bofill}, \citenamefont {Sun}, \citenamefont {Kanne}, \citenamefont {Nyg\aa{}rd}, \citenamefont {Kjaergaard},\ and\ \citenamefont {Fatemi}}]{lu2025}%
  \BibitemOpen
  \bibfield  {author} {\bibinfo {author} {\bibfnamefont {H.}~\bibnamefont {Lu}}, \bibinfo {author} {\bibfnamefont {D.~F.}\ \bibnamefont {Bofill}}, \bibinfo {author} {\bibfnamefont {Z.}~\bibnamefont {Sun}}, \bibinfo {author} {\bibfnamefont {T.}~\bibnamefont {Kanne}}, \bibinfo {author} {\bibfnamefont {J.}~\bibnamefont {Nyg\aa{}rd}}, \bibinfo {author} {\bibfnamefont {M.}~\bibnamefont {Kjaergaard}},\ and\ \bibinfo {author} {\bibfnamefont {V.}~\bibnamefont {Fatemi}},\ }\bibfield  {title} {\bibinfo {title} {Andreev spin relaxation time in a shadow-evaporated {InAs} weak link},\ }\href {https://doi.org/10.1103/v3lq-t5z8} {\bibfield  {journal} {\bibinfo  {journal} {Phys. Rev. Appl.}\ }\textbf {\bibinfo {volume} {24}},\ \bibinfo {pages} {024046} (\bibinfo {year} {2025}{\natexlab{a}})}\BibitemShut {NoStop}%
\bibitem [{\citenamefont {Hoffman}\ \emph {et~al.}(2025)\citenamefont {Hoffman}, \citenamefont {Hays}, \citenamefont {Serniak}, \citenamefont {Hazard},\ and\ \citenamefont {Tahan}}]{hoffman2025}%
  \BibitemOpen
  \bibfield  {author} {\bibinfo {author} {\bibfnamefont {S.}~\bibnamefont {Hoffman}}, \bibinfo {author} {\bibfnamefont {M.}~\bibnamefont {Hays}}, \bibinfo {author} {\bibfnamefont {K.}~\bibnamefont {Serniak}}, \bibinfo {author} {\bibfnamefont {T.}~\bibnamefont {Hazard}},\ and\ \bibinfo {author} {\bibfnamefont {C.}~\bibnamefont {Tahan}},\ }\bibfield  {title} {\bibinfo {title} {Decoherence in {Andreev} spin qubits},\ }\href {https://doi.org/10.1103/PhysRevB.111.045304} {\bibfield  {journal} {\bibinfo  {journal} {Phys. Rev. B}\ }\textbf {\bibinfo {volume} {111}},\ \bibinfo {pages} {045304} (\bibinfo {year} {2025})}\BibitemShut {NoStop}%
\bibitem [{\citenamefont {Stano}\ and\ \citenamefont {Loss}(2022)}]{stano2022}%
  \BibitemOpen
  \bibfield  {author} {\bibinfo {author} {\bibfnamefont {P.}~\bibnamefont {Stano}}\ and\ \bibinfo {author} {\bibfnamefont {D.}~\bibnamefont {Loss}},\ }\bibfield  {title} {\bibinfo {title} {Review of performance metrics of spin qubits in gated semiconducting nanostructures},\ }\href {https://doi.org/https://doi.org/10.1038/s42254-022-00484-w} {\bibfield  {journal} {\bibinfo  {journal} {Nature Reviews Physics}\ }\textbf {\bibinfo {volume} {4}},\ \bibinfo {pages} {672} (\bibinfo {year} {2022})}\BibitemShut {NoStop}%
\bibitem [{\citenamefont {Burkard}\ \emph {et~al.}(2023)\citenamefont {Burkard}, \citenamefont {Ladd}, \citenamefont {Pan}, \citenamefont {Nichol},\ and\ \citenamefont {Petta}}]{burkard2023}%
  \BibitemOpen
  \bibfield  {author} {\bibinfo {author} {\bibfnamefont {G.}~\bibnamefont {Burkard}}, \bibinfo {author} {\bibfnamefont {T.~D.}\ \bibnamefont {Ladd}}, \bibinfo {author} {\bibfnamefont {A.}~\bibnamefont {Pan}}, \bibinfo {author} {\bibfnamefont {J.~M.}\ \bibnamefont {Nichol}},\ and\ \bibinfo {author} {\bibfnamefont {J.~R.}\ \bibnamefont {Petta}},\ }\bibfield  {title} {\bibinfo {title} {Semiconductor spin qubits},\ }\href {https://doi.org/10.1103/RevModPhys.95.025003} {\bibfield  {journal} {\bibinfo  {journal} {Rev. Mod. Phys.}\ }\textbf {\bibinfo {volume} {95}},\ \bibinfo {pages} {025003} (\bibinfo {year} {2023})}\BibitemShut {NoStop}%
\bibitem [{\citenamefont {Sigillito}\ \emph {et~al.}(2015)\citenamefont {Sigillito}, \citenamefont {Jock}, \citenamefont {Tyryshkin}, \citenamefont {Beeman}, \citenamefont {Haller}, \citenamefont {Itoh},\ and\ \citenamefont {Lyon}}]{sigillito2015}%
  \BibitemOpen
  \bibfield  {author} {\bibinfo {author} {\bibfnamefont {A.~J.}\ \bibnamefont {Sigillito}}, \bibinfo {author} {\bibfnamefont {R.~M.}\ \bibnamefont {Jock}}, \bibinfo {author} {\bibfnamefont {A.~M.}\ \bibnamefont {Tyryshkin}}, \bibinfo {author} {\bibfnamefont {J.~W.}\ \bibnamefont {Beeman}}, \bibinfo {author} {\bibfnamefont {E.~E.}\ \bibnamefont {Haller}}, \bibinfo {author} {\bibfnamefont {K.~M.}\ \bibnamefont {Itoh}},\ and\ \bibinfo {author} {\bibfnamefont {S.~A.}\ \bibnamefont {Lyon}},\ }\bibfield  {title} {\bibinfo {title} {Electron spin coherence of shallow donors in natural and isotopically enriched germanium},\ }\href {https://doi.org/10.1103/PhysRevLett.115.247601} {\bibfield  {journal} {\bibinfo  {journal} {Phys. Rev. Lett.}\ }\textbf {\bibinfo {volume} {115}},\ \bibinfo {pages} {247601} (\bibinfo {year} {2015})}\BibitemShut {NoStop}%
\bibitem [{\citenamefont {Lakic}\ \emph {et~al.}(2025)\citenamefont {Lakic}, \citenamefont {Lawrie}, \citenamefont {van Driel}, \citenamefont {Stehouwer}, \citenamefont {Su}, \citenamefont {Veldhorst}, \citenamefont {Scappucci}, \citenamefont {Kuemmeth},\ and\ \citenamefont {Chatterjee}}]{lakic2025}%
  \BibitemOpen
  \bibfield  {author} {\bibinfo {author} {\bibfnamefont {L.}~\bibnamefont {Lakic}}, \bibinfo {author} {\bibfnamefont {W.~I.}\ \bibnamefont {Lawrie}}, \bibinfo {author} {\bibfnamefont {D.}~\bibnamefont {van Driel}}, \bibinfo {author} {\bibfnamefont {L.~E.}\ \bibnamefont {Stehouwer}}, \bibinfo {author} {\bibfnamefont {Y.}~\bibnamefont {Su}}, \bibinfo {author} {\bibfnamefont {M.}~\bibnamefont {Veldhorst}}, \bibinfo {author} {\bibfnamefont {G.}~\bibnamefont {Scappucci}}, \bibinfo {author} {\bibfnamefont {F.}~\bibnamefont {Kuemmeth}},\ and\ \bibinfo {author} {\bibfnamefont {A.}~\bibnamefont {Chatterjee}},\ }\bibfield  {title} {\bibinfo {title} {A quantum dot in germanium proximitized by a superconductor},\ }\href {https://doi.org/https://doi.org/10.1038/s41563-024-02095-5} {\bibfield  {journal} {\bibinfo  {journal} {Nature Materials}\ }\textbf {\bibinfo {volume} {24}},\ \bibinfo {pages} {552} (\bibinfo {year} {2025})}\BibitemShut {NoStop}%
\bibitem [{\citenamefont {Hoffman}\ and\ \citenamefont {Tahan}(2025)}]{hoffman2025ge}%
  \BibitemOpen
  \bibfield  {author} {\bibinfo {author} {\bibfnamefont {S.}~\bibnamefont {Hoffman}}\ and\ \bibinfo {author} {\bibfnamefont {C.}~\bibnamefont {Tahan}},\ }\href {https://arxiv.org/abs/2506.13988} {\bibinfo {title} {Resolving {Andreev spin qubits} in germanium-based {Josephson} junctions}} (\bibinfo {year} {2025}),\ \Eprint {https://arxiv.org/abs/2506.13988} {arXiv:2506.13988 [cond-mat.mes-hall]} \BibitemShut {NoStop}%
\bibitem [{\citenamefont {Gyenis}\ \emph {et~al.}(2021)\citenamefont {Gyenis}, \citenamefont {Di~Paolo}, \citenamefont {Koch}, \citenamefont {Blais}, \citenamefont {Houck},\ and\ \citenamefont {Schuster}}]{gyenis2021}%
  \BibitemOpen
  \bibfield  {author} {\bibinfo {author} {\bibfnamefont {A.}~\bibnamefont {Gyenis}}, \bibinfo {author} {\bibfnamefont {A.}~\bibnamefont {Di~Paolo}}, \bibinfo {author} {\bibfnamefont {J.}~\bibnamefont {Koch}}, \bibinfo {author} {\bibfnamefont {A.}~\bibnamefont {Blais}}, \bibinfo {author} {\bibfnamefont {A.~A.}\ \bibnamefont {Houck}},\ and\ \bibinfo {author} {\bibfnamefont {D.~I.}\ \bibnamefont {Schuster}},\ }\bibfield  {title} {\bibinfo {title} {Moving beyond the transmon: {Noise}-protected superconducting quantum circuits},\ }\href {https://doi.org/10.1103/PRXQuantum.2.030101} {\bibfield  {journal} {\bibinfo  {journal} {PRX Quantum}\ }\textbf {\bibinfo {volume} {2}},\ \bibinfo {pages} {030101} (\bibinfo {year} {2021})}\BibitemShut {NoStop}%
\bibitem [{\citenamefont {Koch}\ \emph {et~al.}(2007)\citenamefont {Koch}, \citenamefont {Yu}, \citenamefont {Gambetta}, \citenamefont {Houck}, \citenamefont {Schuster}, \citenamefont {Majer}, \citenamefont {Blais}, \citenamefont {Devoret}, \citenamefont {Girvin},\ and\ \citenamefont {Schoelkopf}}]{koch2007}%
  \BibitemOpen
  \bibfield  {author} {\bibinfo {author} {\bibfnamefont {J.}~\bibnamefont {Koch}}, \bibinfo {author} {\bibfnamefont {T.~M.}\ \bibnamefont {Yu}}, \bibinfo {author} {\bibfnamefont {J.}~\bibnamefont {Gambetta}}, \bibinfo {author} {\bibfnamefont {A.~A.}\ \bibnamefont {Houck}}, \bibinfo {author} {\bibfnamefont {D.~I.}\ \bibnamefont {Schuster}}, \bibinfo {author} {\bibfnamefont {J.}~\bibnamefont {Majer}}, \bibinfo {author} {\bibfnamefont {A.}~\bibnamefont {Blais}}, \bibinfo {author} {\bibfnamefont {M.~H.}\ \bibnamefont {Devoret}}, \bibinfo {author} {\bibfnamefont {S.~M.}\ \bibnamefont {Girvin}},\ and\ \bibinfo {author} {\bibfnamefont {R.~J.}\ \bibnamefont {Schoelkopf}},\ }\bibfield  {title} {\bibinfo {title} {Charge-insensitive qubit design derived from the {Cooper} pair box},\ }\href {https://doi.org/10.1103/PhysRevA.76.042319} {\bibfield  {journal} {\bibinfo  {journal} {Phys. Rev. A}\ }\textbf {\bibinfo {volume} {76}},\ \bibinfo {pages} {042319} (\bibinfo {year} {2007})}\BibitemShut {NoStop}%
\bibitem [{\citenamefont {Bransden}\ and\ \citenamefont {Joachain}(2003)}]{bransden2003}%
  \BibitemOpen
  \bibfield  {author} {\bibinfo {author} {\bibfnamefont {B.~H.}\ \bibnamefont {Bransden}}\ and\ \bibinfo {author} {\bibfnamefont {C.~J.}\ \bibnamefont {Joachain}},\ }\href@noop {} {\emph {\bibinfo {title} {Physics of {Atoms} and {Molecules}}}}\ (\bibinfo  {publisher} {Pearson Education},\ \bibinfo {year} {2003})\BibitemShut {NoStop}%
\bibitem [{\citenamefont {Koch}\ and\ \citenamefont {von Oppen}(2005)}]{koch2005}%
  \BibitemOpen
  \bibfield  {author} {\bibinfo {author} {\bibfnamefont {J.}~\bibnamefont {Koch}}\ and\ \bibinfo {author} {\bibfnamefont {F.}~\bibnamefont {von Oppen}},\ }\bibfield  {title} {\bibinfo {title} {{Franck-Condon} blockade and giant {Fano} factors in transport through single molecules},\ }\href {https://doi.org/10.1103/PhysRevLett.94.206804} {\bibfield  {journal} {\bibinfo  {journal} {Phys. Rev. Lett.}\ }\textbf {\bibinfo {volume} {94}},\ \bibinfo {pages} {206804} (\bibinfo {year} {2005})}\BibitemShut {NoStop}%
\bibitem [{\citenamefont {Leturcq}\ \emph {et~al.}(2009)\citenamefont {Leturcq}, \citenamefont {Stampfer}, \citenamefont {Inderbitzin}, \citenamefont {Durrer}, \citenamefont {Hierold}, \citenamefont {Mariani}, \citenamefont {Schultz}, \citenamefont {Von~Oppen},\ and\ \citenamefont {Ensslin}}]{leturcq2009}%
  \BibitemOpen
  \bibfield  {author} {\bibinfo {author} {\bibfnamefont {R.}~\bibnamefont {Leturcq}}, \bibinfo {author} {\bibfnamefont {C.}~\bibnamefont {Stampfer}}, \bibinfo {author} {\bibfnamefont {K.}~\bibnamefont {Inderbitzin}}, \bibinfo {author} {\bibfnamefont {L.}~\bibnamefont {Durrer}}, \bibinfo {author} {\bibfnamefont {C.}~\bibnamefont {Hierold}}, \bibinfo {author} {\bibfnamefont {E.}~\bibnamefont {Mariani}}, \bibinfo {author} {\bibfnamefont {M.~G.}\ \bibnamefont {Schultz}}, \bibinfo {author} {\bibfnamefont {F.}~\bibnamefont {Von~Oppen}},\ and\ \bibinfo {author} {\bibfnamefont {K.}~\bibnamefont {Ensslin}},\ }\bibfield  {title} {\bibinfo {title} {{Franck--Condon} blockade in suspended carbon nanotube quantum dots},\ }\href {https://doi.org/https://doi.org/10.1038/nphys1234} {\bibfield  {journal} {\bibinfo  {journal} {Nature Physics}\ }\textbf {\bibinfo {volume} {5}},\ \bibinfo {pages} {327} (\bibinfo {year} {2009})}\BibitemShut {NoStop}%
\bibitem [{\citenamefont {Bargerbos}\ \emph {et~al.}(2023)\citenamefont {Bargerbos}, \citenamefont {Pita-Vidal}, \citenamefont {\ifmmode~\check{Z}\else \v{Z}\fi{}itko}, \citenamefont {Splitthoff}, \citenamefont {Gr\"unhaupt}, \citenamefont {Wesdorp}, \citenamefont {Liu}, \citenamefont {Kouwenhoven}, \citenamefont {Aguado}, \citenamefont {Andersen}, \citenamefont {Kou},\ and\ \citenamefont {van Heck}}]{bargerbos2023}%
  \BibitemOpen
  \bibfield  {author} {\bibinfo {author} {\bibfnamefont {A.}~\bibnamefont {Bargerbos}}, \bibinfo {author} {\bibfnamefont {M.}~\bibnamefont {Pita-Vidal}}, \bibinfo {author} {\bibfnamefont {R.}~\bibnamefont {\ifmmode~\check{Z}\else \v{Z}\fi{}itko}}, \bibinfo {author} {\bibfnamefont {L.~J.}\ \bibnamefont {Splitthoff}}, \bibinfo {author} {\bibfnamefont {L.}~\bibnamefont {Gr\"unhaupt}}, \bibinfo {author} {\bibfnamefont {J.~J.}\ \bibnamefont {Wesdorp}}, \bibinfo {author} {\bibfnamefont {Y.}~\bibnamefont {Liu}}, \bibinfo {author} {\bibfnamefont {L.~P.}\ \bibnamefont {Kouwenhoven}}, \bibinfo {author} {\bibfnamefont {R.}~\bibnamefont {Aguado}}, \bibinfo {author} {\bibfnamefont {C.~K.}\ \bibnamefont {Andersen}}, \bibinfo {author} {\bibfnamefont {A.}~\bibnamefont {Kou}},\ and\ \bibinfo {author} {\bibfnamefont {B.}~\bibnamefont {van Heck}},\ }\bibfield  {title} {\bibinfo {title} {Spectroscopy of spin-split {Andreev} levels in a quantum dot with superconducting leads},\ }\href
  {https://doi.org/10.1103/PhysRevLett.131.097001} {\bibfield  {journal} {\bibinfo  {journal} {Phys. Rev. Lett.}\ }\textbf {\bibinfo {volume} {131}},\ \bibinfo {pages} {097001} (\bibinfo {year} {2023})}\BibitemShut {NoStop}%
\bibitem [{\citenamefont {Erlingsson}\ and\ \citenamefont {Nazarov}(2002)}]{erlingsson2002}%
  \BibitemOpen
  \bibfield  {author} {\bibinfo {author} {\bibfnamefont {S.~I.}\ \bibnamefont {Erlingsson}}\ and\ \bibinfo {author} {\bibfnamefont {Y.~V.}\ \bibnamefont {Nazarov}},\ }\bibfield  {title} {\bibinfo {title} {Hyperfine-mediated transitions between a {Zeeman} split doublet in {GaAs} quantum dots: {The} role of the internal field},\ }\href {https://doi.org/10.1103/PhysRevB.66.155327} {\bibfield  {journal} {\bibinfo  {journal} {Phys. Rev. B}\ }\textbf {\bibinfo {volume} {66}},\ \bibinfo {pages} {155327} (\bibinfo {year} {2002})}\BibitemShut {NoStop}%
\bibitem [{\citenamefont {Glazman}\ and\ \citenamefont {Catelani}(2021)}]{glazman2021}%
  \BibitemOpen
  \bibfield  {author} {\bibinfo {author} {\bibfnamefont {L.~I.}\ \bibnamefont {Glazman}}\ and\ \bibinfo {author} {\bibfnamefont {G.}~\bibnamefont {Catelani}},\ }\bibfield  {title} {\bibinfo {title} {{Bogoliubov quasiparticles in superconducting qubits}},\ }\href {https://doi.org/10.21468/SciPostPhysLectNotes.31} {\bibfield  {journal} {\bibinfo  {journal} {SciPost Phys. Lect. Notes}\ ,\ \bibinfo {pages} {31}} (\bibinfo {year} {2021})}\BibitemShut {NoStop}%
\bibitem [{\citenamefont {Frattini}\ \emph {et~al.}(2024)\citenamefont {Frattini}, \citenamefont {Corti\~nas}, \citenamefont {Venkatraman}, \citenamefont {Xiao}, \citenamefont {Su}, \citenamefont {Lei}, \citenamefont {Chapman}, \citenamefont {Joshi}, \citenamefont {Girvin}, \citenamefont {Schoelkopf}, \citenamefont {Puri},\ and\ \citenamefont {Devoret}}]{Frattini2024}%
  \BibitemOpen
  \bibfield  {author} {\bibinfo {author} {\bibfnamefont {N.~E.}\ \bibnamefont {Frattini}}, \bibinfo {author} {\bibfnamefont {R.~G.}\ \bibnamefont {Corti\~nas}}, \bibinfo {author} {\bibfnamefont {J.}~\bibnamefont {Venkatraman}}, \bibinfo {author} {\bibfnamefont {X.}~\bibnamefont {Xiao}}, \bibinfo {author} {\bibfnamefont {Q.}~\bibnamefont {Su}}, \bibinfo {author} {\bibfnamefont {C.~U.}\ \bibnamefont {Lei}}, \bibinfo {author} {\bibfnamefont {B.~J.}\ \bibnamefont {Chapman}}, \bibinfo {author} {\bibfnamefont {V.~R.}\ \bibnamefont {Joshi}}, \bibinfo {author} {\bibfnamefont {S.~M.}\ \bibnamefont {Girvin}}, \bibinfo {author} {\bibfnamefont {R.~J.}\ \bibnamefont {Schoelkopf}}, \bibinfo {author} {\bibfnamefont {S.}~\bibnamefont {Puri}},\ and\ \bibinfo {author} {\bibfnamefont {M.~H.}\ \bibnamefont {Devoret}},\ }\bibfield  {title} {\bibinfo {title} {Observation of pairwise level degeneracies and the quantum regime of the {Arrhenius} law in a double-well parametric oscillator},\ }\href
  {https://doi.org/10.1103/PhysRevX.14.031040} {\bibfield  {journal} {\bibinfo  {journal} {Phys. Rev. X}\ }\textbf {\bibinfo {volume} {14}},\ \bibinfo {pages} {031040} (\bibinfo {year} {2024})}\BibitemShut {NoStop}%
\bibitem [{\citenamefont {Nowack}\ \emph {et~al.}(2007)\citenamefont {Nowack}, \citenamefont {Koppens}, \citenamefont {Nazarov},\ and\ \citenamefont {Vandersypen}}]{nowack2007}%
  \BibitemOpen
  \bibfield  {author} {\bibinfo {author} {\bibfnamefont {K.~C.}\ \bibnamefont {Nowack}}, \bibinfo {author} {\bibfnamefont {F.}~\bibnamefont {Koppens}}, \bibinfo {author} {\bibfnamefont {Y.~V.}\ \bibnamefont {Nazarov}},\ and\ \bibinfo {author} {\bibfnamefont {L.}~\bibnamefont {Vandersypen}},\ }\bibfield  {title} {\bibinfo {title} {Coherent control of a single electron spin with electric fields},\ }\href {https://doi.org/https://doi.org/10.1126/science.1148092} {\bibfield  {journal} {\bibinfo  {journal} {Science}\ }\textbf {\bibinfo {volume} {318}},\ \bibinfo {pages} {1430} (\bibinfo {year} {2007})}\BibitemShut {NoStop}%
\bibitem [{\citenamefont {Nadj-Perge}\ \emph {et~al.}(2010)\citenamefont {Nadj-Perge}, \citenamefont {Frolov}, \citenamefont {Bakkers},\ and\ \citenamefont {Kouwenhoven}}]{nadj2010}%
  \BibitemOpen
  \bibfield  {author} {\bibinfo {author} {\bibfnamefont {S.}~\bibnamefont {Nadj-Perge}}, \bibinfo {author} {\bibfnamefont {S.~M.}\ \bibnamefont {Frolov}}, \bibinfo {author} {\bibfnamefont {E.~P. A.~M.}\ \bibnamefont {Bakkers}},\ and\ \bibinfo {author} {\bibfnamefont {L.~P.}\ \bibnamefont {Kouwenhoven}},\ }\bibfield  {title} {\bibinfo {title} {Spin--orbit qubit in a semiconductor nanowire},\ }\href {https://doi.org/https://doi.org/10.1038/nature09682} {\bibfield  {journal} {\bibinfo  {journal} {Nature}\ }\textbf {\bibinfo {volume} {468}},\ \bibinfo {pages} {1084} (\bibinfo {year} {2010})}\BibitemShut {NoStop}%
\bibitem [{\citenamefont {Golovach}\ \emph {et~al.}(2006)\citenamefont {Golovach}, \citenamefont {Borhani},\ and\ \citenamefont {Loss}}]{golovach2006}%
  \BibitemOpen
  \bibfield  {author} {\bibinfo {author} {\bibfnamefont {V.~N.}\ \bibnamefont {Golovach}}, \bibinfo {author} {\bibfnamefont {M.}~\bibnamefont {Borhani}},\ and\ \bibinfo {author} {\bibfnamefont {D.}~\bibnamefont {Loss}},\ }\bibfield  {title} {\bibinfo {title} {Electric-dipole-induced spin resonance in quantum dots},\ }\href {https://doi.org/10.1103/PhysRevB.74.165319} {\bibfield  {journal} {\bibinfo  {journal} {Phys. Rev. B}\ }\textbf {\bibinfo {volume} {74}},\ \bibinfo {pages} {165319} (\bibinfo {year} {2006})}\BibitemShut {NoStop}%
\bibitem [{\citenamefont {Kurilovich}\ \emph {et~al.}(2025)\citenamefont {Kurilovich}, \citenamefont {Connolly}, \citenamefont {Bøttcher}, \citenamefont {Weiss}, \citenamefont {Hazra}, \citenamefont {Joshi}, \citenamefont {Ding}, \citenamefont {Nho}, \citenamefont {Diamond}, \citenamefont {Kurilovich}, \citenamefont {Dai}, \citenamefont {Fatemi}, \citenamefont {Frunzio}, \citenamefont {Glazman},\ and\ \citenamefont {Devoret}}]{kurilovich2025}%
  \BibitemOpen
  \bibfield  {author} {\bibinfo {author} {\bibfnamefont {P.~D.}\ \bibnamefont {Kurilovich}}, \bibinfo {author} {\bibfnamefont {T.}~\bibnamefont {Connolly}}, \bibinfo {author} {\bibfnamefont {C.~G.~L.}\ \bibnamefont {Bøttcher}}, \bibinfo {author} {\bibfnamefont {D.~K.}\ \bibnamefont {Weiss}}, \bibinfo {author} {\bibfnamefont {S.}~\bibnamefont {Hazra}}, \bibinfo {author} {\bibfnamefont {V.~R.}\ \bibnamefont {Joshi}}, \bibinfo {author} {\bibfnamefont {A.~Z.}\ \bibnamefont {Ding}}, \bibinfo {author} {\bibfnamefont {H.}~\bibnamefont {Nho}}, \bibinfo {author} {\bibfnamefont {S.}~\bibnamefont {Diamond}}, \bibinfo {author} {\bibfnamefont {V.~D.}\ \bibnamefont {Kurilovich}}, \bibinfo {author} {\bibfnamefont {W.}~\bibnamefont {Dai}}, \bibinfo {author} {\bibfnamefont {V.}~\bibnamefont {Fatemi}}, \bibinfo {author} {\bibfnamefont {L.}~\bibnamefont {Frunzio}}, \bibinfo {author} {\bibfnamefont {L.~I.}\ \bibnamefont {Glazman}},\ and\ \bibinfo {author} {\bibfnamefont {M.~H.}\ \bibnamefont {Devoret}},\ }\href
  {https://arxiv.org/abs/2501.09161} {\bibinfo {title} {High-frequency readout free from transmon multi-excitation resonances}} (\bibinfo {year} {2025}),\ \Eprint {https://arxiv.org/abs/2501.09161} {arXiv:2501.09161 [quant-ph]} \BibitemShut {NoStop}%
\bibitem [{Note1()}]{Note1}%
  \BibitemOpen
  \bibinfo {note} {See e.g. the Supplemental Materiak of Ref.~\cite {bargerbos2023} for a demontration of Andreev spin readout via a resonator.}\BibitemShut {Stop}%
\bibitem [{\citenamefont {Pistolesi}\ \emph {et~al.}(2021)\citenamefont {Pistolesi}, \citenamefont {Cleland},\ and\ \citenamefont {Bachtold}}]{pistolesi2021}%
  \BibitemOpen
  \bibfield  {author} {\bibinfo {author} {\bibfnamefont {F.}~\bibnamefont {Pistolesi}}, \bibinfo {author} {\bibfnamefont {A.~N.}\ \bibnamefont {Cleland}},\ and\ \bibinfo {author} {\bibfnamefont {A.}~\bibnamefont {Bachtold}},\ }\bibfield  {title} {\bibinfo {title} {Proposal for a nanomechanical qubit},\ }\href {https://doi.org/10.1103/PhysRevX.11.031027} {\bibfield  {journal} {\bibinfo  {journal} {Phys. Rev. X}\ }\textbf {\bibinfo {volume} {11}},\ \bibinfo {pages} {031027} (\bibinfo {year} {2021})}\BibitemShut {NoStop}%
\bibitem [{\citenamefont {Aliferis}\ and\ \citenamefont {Preskill}(2008)}]{aliferis2008}%
  \BibitemOpen
  \bibfield  {author} {\bibinfo {author} {\bibfnamefont {P.}~\bibnamefont {Aliferis}}\ and\ \bibinfo {author} {\bibfnamefont {J.}~\bibnamefont {Preskill}},\ }\bibfield  {title} {\bibinfo {title} {Fault-tolerant quantum computation against biased noise},\ }\href {https://doi.org/10.1103/PhysRevA.78.052331} {\bibfield  {journal} {\bibinfo  {journal} {Phys. Rev. A}\ }\textbf {\bibinfo {volume} {78}},\ \bibinfo {pages} {052331} (\bibinfo {year} {2008})}\BibitemShut {NoStop}%
\bibitem [{\citenamefont {Tuckett}\ \emph {et~al.}(2018)\citenamefont {Tuckett}, \citenamefont {Bartlett},\ and\ \citenamefont {Flammia}}]{tuckett2018}%
  \BibitemOpen
  \bibfield  {author} {\bibinfo {author} {\bibfnamefont {D.~K.}\ \bibnamefont {Tuckett}}, \bibinfo {author} {\bibfnamefont {S.~D.}\ \bibnamefont {Bartlett}},\ and\ \bibinfo {author} {\bibfnamefont {S.~T.}\ \bibnamefont {Flammia}},\ }\bibfield  {title} {\bibinfo {title} {Ultrahigh error threshold for surface codes with biased noise},\ }\href {https://doi.org/10.1103/PhysRevLett.120.050505} {\bibfield  {journal} {\bibinfo  {journal} {Phys. Rev. Lett.}\ }\textbf {\bibinfo {volume} {120}},\ \bibinfo {pages} {050505} (\bibinfo {year} {2018})}\BibitemShut {NoStop}%
\bibitem [{\citenamefont {Tuckett}\ \emph {et~al.}(2019)\citenamefont {Tuckett}, \citenamefont {Darmawan}, \citenamefont {Chubb}, \citenamefont {Bravyi}, \citenamefont {Bartlett},\ and\ \citenamefont {Flammia}}]{tuckett2019}%
  \BibitemOpen
  \bibfield  {author} {\bibinfo {author} {\bibfnamefont {D.~K.}\ \bibnamefont {Tuckett}}, \bibinfo {author} {\bibfnamefont {A.~S.}\ \bibnamefont {Darmawan}}, \bibinfo {author} {\bibfnamefont {C.~T.}\ \bibnamefont {Chubb}}, \bibinfo {author} {\bibfnamefont {S.}~\bibnamefont {Bravyi}}, \bibinfo {author} {\bibfnamefont {S.~D.}\ \bibnamefont {Bartlett}},\ and\ \bibinfo {author} {\bibfnamefont {S.~T.}\ \bibnamefont {Flammia}},\ }\bibfield  {title} {\bibinfo {title} {Tailoring surface codes for highly biased noise},\ }\href {https://doi.org/10.1103/PhysRevX.9.041031} {\bibfield  {journal} {\bibinfo  {journal} {Phys. Rev. X}\ }\textbf {\bibinfo {volume} {9}},\ \bibinfo {pages} {041031} (\bibinfo {year} {2019})}\BibitemShut {NoStop}%
\bibitem [{\citenamefont {R{\'e}glade}\ \emph {et~al.}(2024)\citenamefont {R{\'e}glade}, \citenamefont {Bocquet}, \citenamefont {Gautier}, \citenamefont {Cohen}, \citenamefont {Marquet}, \citenamefont {Albertinale}, \citenamefont {Pankratova}, \citenamefont {Hall{\'e}n}, \citenamefont {Rautschke}, \citenamefont {Sellem} \emph {et~al.}}]{reglade2024quantum}%
  \BibitemOpen
  \bibfield  {author} {\bibinfo {author} {\bibfnamefont {U.}~\bibnamefont {R{\'e}glade}}, \bibinfo {author} {\bibfnamefont {A.}~\bibnamefont {Bocquet}}, \bibinfo {author} {\bibfnamefont {R.}~\bibnamefont {Gautier}}, \bibinfo {author} {\bibfnamefont {J.}~\bibnamefont {Cohen}}, \bibinfo {author} {\bibfnamefont {A.}~\bibnamefont {Marquet}}, \bibinfo {author} {\bibfnamefont {E.}~\bibnamefont {Albertinale}}, \bibinfo {author} {\bibfnamefont {N.}~\bibnamefont {Pankratova}}, \bibinfo {author} {\bibfnamefont {M.}~\bibnamefont {Hall{\'e}n}}, \bibinfo {author} {\bibfnamefont {F.}~\bibnamefont {Rautschke}}, \bibinfo {author} {\bibfnamefont {L.-A.}\ \bibnamefont {Sellem}}, \emph {et~al.},\ }\bibfield  {title} {\bibinfo {title} {Quantum control of a cat qubit with bit-flip times exceeding ten seconds},\ }\href {https://doi.org/https://doi.org/10.1038/s41586-024-07294-3} {\bibfield  {journal} {\bibinfo  {journal} {Nature}\ }\textbf {\bibinfo {volume} {629}},\ \bibinfo {pages} {778} (\bibinfo {year} {2024})}\BibitemShut
  {NoStop}%
\bibitem [{\citenamefont {Lu}\ \emph {et~al.}(2025{\natexlab{b}})\citenamefont {Lu}, \citenamefont {Day}, \citenamefont {Akhmerov}, \citenamefont {van Heck},\ and\ \citenamefont {Fatemi}}]{lu2025kramers}%
  \BibitemOpen
  \bibfield  {author} {\bibinfo {author} {\bibfnamefont {H.}~\bibnamefont {Lu}}, \bibinfo {author} {\bibfnamefont {I.~A.}\ \bibnamefont {Day}}, \bibinfo {author} {\bibfnamefont {A.~R.}\ \bibnamefont {Akhmerov}}, \bibinfo {author} {\bibfnamefont {B.}~\bibnamefont {van Heck}},\ and\ \bibinfo {author} {\bibfnamefont {V.}~\bibnamefont {Fatemi}},\ }\href {https://arxiv.org/abs/2412.16116} {\bibinfo {title} {Kramers-protected hardware-efficient error correction with {Andreev} spin qubits}} (\bibinfo {year} {2025}{\natexlab{b}}),\ \Eprint {https://arxiv.org/abs/2412.16116} {arXiv:2412.16116 [quant-ph]} \BibitemShut {NoStop}%
\bibitem [{\citenamefont {Manucharyan}\ \emph {et~al.}(2009)\citenamefont {Manucharyan}, \citenamefont {Koch}, \citenamefont {Glazman},\ and\ \citenamefont {Devoret}}]{manucharyan2009}%
  \BibitemOpen
  \bibfield  {author} {\bibinfo {author} {\bibfnamefont {V.~E.}\ \bibnamefont {Manucharyan}}, \bibinfo {author} {\bibfnamefont {J.}~\bibnamefont {Koch}}, \bibinfo {author} {\bibfnamefont {L.~I.}\ \bibnamefont {Glazman}},\ and\ \bibinfo {author} {\bibfnamefont {M.~H.}\ \bibnamefont {Devoret}},\ }\bibfield  {title} {\bibinfo {title} {Fluxonium: {Single} {Cooper}-pair circuit free of charge offsets},\ }\href {https://doi.org/10.1126/science.1175552} {\bibfield  {journal} {\bibinfo  {journal} {Science}\ }\textbf {\bibinfo {volume} {326}},\ \bibinfo {pages} {113} (\bibinfo {year} {2009})}\BibitemShut {NoStop}%
\bibitem [{\citenamefont {Olmo}\ \emph {et~al.}()\citenamefont {Olmo}, \citenamefont {Matute-Cañadas}, \citenamefont {Souto}, \citenamefont {Levy~Yeyati},\ and\ \citenamefont {Aguado}}]{olmoinprep}%
  \BibitemOpen
  \bibfield  {author} {\bibinfo {author} {\bibfnamefont {J.~L.}\ \bibnamefont {Olmo}}, \bibinfo {author} {\bibfnamefont {F.~J.}\ \bibnamefont {Matute-Cañadas}}, \bibinfo {author} {\bibfnamefont {R.~S.}\ \bibnamefont {Souto}}, \bibinfo {author} {\bibfnamefont {A.}~\bibnamefont {Levy~Yeyati}},\ and\ \bibinfo {author} {\bibfnamefont {R.}~\bibnamefont {Aguado}},\ }\bibinfo {note} {in preparation}\BibitemShut {NoStop}%
\bibitem [{\citenamefont {van Heck}\ \emph {et~al.}(2014)\citenamefont {van Heck}, \citenamefont {Mi},\ and\ \citenamefont {Akhmerov}}]{heck2014}%
  \BibitemOpen
  \bibfield  {author} {\bibinfo {author} {\bibfnamefont {B.}~\bibnamefont {van Heck}}, \bibinfo {author} {\bibfnamefont {S.}~\bibnamefont {Mi}},\ and\ \bibinfo {author} {\bibfnamefont {A.~R.}\ \bibnamefont {Akhmerov}},\ }\bibfield  {title} {\bibinfo {title} {Single fermion manipulation via superconducting phase differences in multiterminal {Josephson} junctions},\ }\href {https://doi.org/10.1103/PhysRevB.90.155450} {\bibfield  {journal} {\bibinfo  {journal} {Phys. Rev. B}\ }\textbf {\bibinfo {volume} {90}},\ \bibinfo {pages} {155450} (\bibinfo {year} {2014})}\BibitemShut {NoStop}%
\bibitem [{\citenamefont {Matute-Ca\~nadas}\ \emph {et~al.}(2024)\citenamefont {Matute-Ca\~nadas}, \citenamefont {Tosi},\ and\ \citenamefont {Yeyati}}]{matute2024}%
  \BibitemOpen
  \bibfield  {author} {\bibinfo {author} {\bibfnamefont {F.}~\bibnamefont {Matute-Ca\~nadas}}, \bibinfo {author} {\bibfnamefont {L.}~\bibnamefont {Tosi}},\ and\ \bibinfo {author} {\bibfnamefont {A.~L.}\ \bibnamefont {Yeyati}},\ }\bibfield  {title} {\bibinfo {title} {Quantum circuits with multiterminal {Josephson-Andreev} junctions},\ }\href {https://doi.org/10.1103/PRXQuantum.5.020340} {\bibfield  {journal} {\bibinfo  {journal} {PRX Quantum}\ }\textbf {\bibinfo {volume} {5}},\ \bibinfo {pages} {020340} (\bibinfo {year} {2024})}\BibitemShut {NoStop}%
\bibitem [{\citenamefont {Piasotski}\ \emph {et~al.}(2024)\citenamefont {Piasotski}, \citenamefont {Svetogorov}, \citenamefont {Belzig},\ and\ \citenamefont {Pletyukhov}}]{piasotski2024}%
  \BibitemOpen
  \bibfield  {author} {\bibinfo {author} {\bibfnamefont {K.}~\bibnamefont {Piasotski}}, \bibinfo {author} {\bibfnamefont {A.}~\bibnamefont {Svetogorov}}, \bibinfo {author} {\bibfnamefont {W.}~\bibnamefont {Belzig}},\ and\ \bibinfo {author} {\bibfnamefont {M.}~\bibnamefont {Pletyukhov}},\ }\href {https://arxiv.org/abs/2411.11155} {\bibinfo {title} {Theory of three-terminal {Andreev} spin qubits}} (\bibinfo {year} {2024}),\ \Eprint {https://arxiv.org/abs/2411.11155} {arXiv:2411.11155 [cond-mat.mes-hall]} \BibitemShut {NoStop}%
\bibitem [{\citenamefont {van Heck}(2025)}]{zenodo}%
  \BibitemOpen
  \bibfield  {author} {\bibinfo {author} {\bibfnamefont {B.}~\bibnamefont {van Heck}},\ }\href {https://doi.org/10.5281/zenodo.15613283} {\bibinfo {title} {Andreev spin qubit protected by {Franck-Condon} blockade}},\ \bibinfo {howpublished} {Dataset on Zenodo} (\bibinfo {year} {2025})\BibitemShut {NoStop}%
\end{thebibliography}%

\end{document}